# Natural Regulation of Energy Flow in a Green Quantum Photocell

Trevor B. Arp[1,2], Yafis Barlas[1,3], Vivek Aji[1], Nathaniel M. Gabor[1,2,3]*

[1]*Department of Physics and Astronomy, University of California, Riverside, CA 92521, USA.*

[2]*Laboratory of Quantum Materials Optoelectronics, University of California, Riverside, CA 92521, USA.*

[3]*Center for Spins and Heat in Nanoscale Electronic Systems (SHINES), University of California, Riverside, CA 92521, USA.*

*Correspondence to: nathaniel.gabor@ucr.edu

**Abstract**

Manipulating the flow of energy in nanoscale and molecular photonic devices is of both fundamental interest and central importance for applications in light harvesting optoelectronics. Under erratic solar irradiance conditions, unregulated power fluctuations in a light harvesting photocell lead to inefficient energy storage in conventional solar cells and potentially fatal oxidative damage in photosynthesis. Here, we show that regulation against these fluctuations arises naturally within a two-channel quantum heat engine photocell, thus enabling the efficient conversion of varying incident solar power into a steady output for photon absorption over a wide range of the solar spectrum at Earth's surface. Remarkably, absorption in the green portion of the spectrum is avoided, as it provides no inherent regulatory benefit. Our findings illuminate a quantum structural origin of regulation, provide a novel optoelectronic design strategy, and may elucidate the link between photoprotection in photosynthesis and the predominance of green plants on Earth.

Thermodynamic heat engines manipulate the flow of energy between two reservoirs, hot and cold, to extract power from the temperature difference[1]. Quantum heat engines (QHEs) operate under the same principle, yet are composed of quantum states, such as discrete electronic energy levels coupled to thermal reservoirs through narrow energy bands[2,3]. QHEs are ubiquitous in solid state physics - important examples include lasers, microcavity-coupled excitons[4-6], Pomeranchuk cooling[7], and adiabatic demagnetization[8,9] - but have recently emerged as a compelling description of biological light-harvesting systems. The observation of coherence effects and wavelike energy transfer within the light harvesting pigment-protein complexes[10-14] has inspired tremendous effort to understand the importance of quantum processes in photosynthesis, particularly within the framework of a quantum photocell[15-19]. By combining molecules or nanoscale semiconductors, individual quantum materials can be linked by an electronic state transition, forming a QHE photocell with one absorbing channel (Fig. 1a). This elementary quantum photocell absorbs a photon from the incident spectral irradiance and converts the photon energy into electronic kinetic energy, thus converting incident photon energy flux into useful low entropy electronic power[15-21].

Energy flowing into a solar cell varies in time corresponding to fluctuations in the incident solar power[22-25]. In conventional photovoltaic technology, fluctuations are suppressed by voltage converters and feedback controllers placed between the solar panel and battery[22]. Switching devices, such as metal-oxide-semiconductor field effect transistors (MOSFETs), modulate the flow of energy in order to provide a regulated voltage for efficient energy storage. In photosynthesis, the suppression of fluctuations is crucial for survival, since excess energy drives severe oxidative damage[23-25]. A hierarchy of photo-protective mechanisms has been identified from macroscopic to microscopic scales[24], where recent measurements suggest that reversible molecular-scale configuration changes regulate energy flow to the reaction center[25-30]. In photovoltaics and photosynthesis, regulation is essential to avoid the accumulation of excess energy, and is achieved by suppressing power fluctuations while matching the input energy flux to the output power demand.

Here, we show that by introducing two absorbing quantum channels, regulation emerges naturally within a QHE photocell. One channel absorbs at a wavelength for which the average input power is high, while the other absorbs at low power. The photocell switches stochastically between high and low power to convert varying incident solar power into steady state output. For the solar power incident at Earth's surface, we evaluated the absorption characteristics that maximize the difference between high and low input power $\Delta u$, a direct measure of solar power variations over which the quantum photocell can regulate. We find that stochastic switching suppresses power fluctuations for absorption over a broad range of the solar spectrum, except near the peak where $\Delta u$ vanishes. Remarkably, the absorption of green light - the most radiant portion of the solar power spectrum per unit wavelength - is avoided, as it provides no regulatory benefit. In addition to informing new design strategies for ultra-efficient biophotonic and optoelectronic devices, the natural regulation process underscores the critical role played by quantum structure in the regulatory function of photosynthetic light harvesting photocells.

**Results**

The two-channel quantum heat engine (2QHE) photocell (Fig. 1b) is composed of two input channels that absorb at similar input energies $E_a$ and $E_b$, while absorbing distinct input powers $u_a < u_b$. The input channels are coupled through electron transfer to the same output ('machine' labeled $M$) at energy $E_M$, and the ratio of input to output energy is comparable for both channels so that neither is energetically favored. The photocell absorbs photons of wavelength $\lambda$ characterized by an incident irradiance spectrum $I(\lambda,T)$, which on Earth is provided by the Sun (a hot thermal reservoir at temperature $T_H \sim 5.8 \times 10^3$ K)[20,21,31]. The absorbed solar power provides an average input energy flux $u_i = \int A_i(\lambda)I(\lambda,T)d\lambda$, where $A_i(\lambda)$ is the absorption spectrum for the $i = a$ or $b$ channel, and $A_i(\lambda)I(\lambda,T)$ is the differential input energy flux. Photon absorption leads to an excited state electron in $|a\rangle$ or $|b\rangle$ that is transferred to the state $|x\rangle$, resulting in energy loss to phonons (a cold reservoir at ambient temperature). We consider the case in which radiative recombination events in the absorbers $a$ and $b$ are rare compared to fast charge transfer to the machine. After charge transfer, which is governed by the transition rates $\gamma_a$ and $\gamma_b$, the excited state energy of the electron that populates the machine $M$ is utilized to generate electronic power $u_M = E_M \Gamma$, where $\Gamma$ is the rate at which energy $E_M$ is produced. To complete the cycle, the low energy electron in $|y\rangle$ returns to the ground state $|g\rangle$ [15-19].

Analyzing the measured solar spectrum at Earth's surface[31], we evaluated $\Delta u = u_b - u_a$ for absorption peaks of width $w$ and peak separation $\Delta\lambda$ as a function of their average, or center, wavelength $\lambda_0$ (Fig. 2a). Figure 2b, our main result, shows that for increasing values of the width $w$, $\Delta u$ vs. $\lambda_0$ transitions from numerous narrow peaks ($w = 5$ nm, blue line Figure 2b, top) to *two* broad peaks (solid green line) and a clear minimum near $\lambda_0 \sim 540$ nm. For the same range of wavelengths within the solar spectrum, we also compared the internal power fluctuations of a two-channel photocell $\sigma_{II}$ to that of a one-channel photocell $\sigma_I$. Figure 2b, bottom shows $\Sigma^2/u_M^2 = (\sigma_I^2 - \sigma_{II}^2)/u_M^2$ vs. $\lambda_0$, which is simply the difference between the squared fluctuations, or variance, of the one- and two-channel photocells. $\Sigma^2/u_M^2$ vs. $\lambda_0$ is always non-negative and exhibits a distinct transition as $w$ increases, from multiple narrow peaks to *two* broad peaks (solid green line) with a single minimum.

Strikingly, as the widths of the absorption peaks approach $w = 35$ nm, $\Delta u$ and $\Sigma^2/u_M^2$ converge to nearly identical peak and minimum values as a function of wavelength. We extracted the values of $\lambda_0$ and $\Delta\lambda$ that simultaneously maximized $\Delta u$ and $\Sigma^2/u_M^2$ to determine the differential input energy flux, $A_a(\lambda)I(\lambda,T)$ and $A_b(\lambda)I(\lambda,T)$ (Fig. 2b, inset), and energy level spacing for two regulating photocells (Fig. 2c). In the following, we examine the wavelength dependence of $\Delta u$ and $\Sigma^2/u_M^2$ by first describing the stochastic energy flow through the photocell, and then assessing the absorption characteristics that mediate intrinsic regulation.

We model the power throughput of the photocell by discretizing time and considering the probabilistic absorption or production of energy at each time step (details of the statistical analysis are presented in Supplementary Section 1). In the steady state limit, energy $E_M$ is produced at every time step. The total energy is then the sum of all events in the past and the long time behavior can be evaluated by the mean and variance at each step. For a single absorbing channel, the photocell switches between the on-state (absorbing $u_b$) and off-state (absorbing nothing) to modulate the input energy flux (Fig. 3a, left). Choosing a probability $p$ that matches the input power distribution to the average output power $pu_b = u_M$, we determined the one-channel photocell variance $\sigma_I^2/u_M^2 = u_b/u_M - 1$. The resulting internal fluctuations $\sigma_I$ are unavoidable.

For two absorbing channels, minimal time is spent in the off-state and the photocell switches between two on-states absorbing high power $u_b$ or low power $u_a$ (Fig. 3a, right), resulting in the minimum variance

$$\sigma_{II}^2/u_M^2 = (u_b/u_M - 1)(1 - u_a/u_M), \qquad (1)$$

where $u_b > u_M > u_a$. The two-channel variance is suppressed below the one-channel photocell variance by the factor $(1 - u_a/u_M)$. For one input with an energy flux larger than the output power ($u_b > u_M$), it is beneficial to have the second input below the output power to suppress internal fluctuations.

To test the validity of the steady state fluctuations (equation (1)), we developed a two-channel mean reversion procedure that simulates realistic conditions, in which the machine outputs energy only when sufficient input energy enters the photocell (See Methods). Figure 3b shows a color map of the resultant two-channel variance $\sigma_{II}^2/u_M^2$ vs. $u_a$ and $u_b$ determined statistically from ~$10^8$ simulated energy flow sequences (calculation details and examples are shown in Supplementary Section 2). We observed that $\sigma_{II}^2/u_M^2$ exhibits a contour of minimum values when the input power nearly matches the output power, i.e. when $u_i/u_M \sim 1$. A horizontal line trace taken from Fig. 3b shows that $\sigma_{II}^2/u_M^2$ decreases linearly as a function of $u_a$ and is fully suppressed when $u_a \sim u_M$ (Fig. 3c). Inversely, the variance increases linearly with $u_b/u_M$ (Fig. 3d, blue line). To compare the simulated behavior to the steady state limit, Figure 3b inset shows $\sigma_{II}^2/u_M^2$ vs. $u_a$ and $u_b$ determined from equation (1). The numerical variance shows excellent agreement with the steady state limit, confirming that fluctuations are suppressed at all times.

Comparison of the energy flow in the one- and two-channel photocells highlights the key advantage of two input channels. For absorption in the blue and red regions of the solar spectrum, power fluctuations of the two-channel photocell are always less than those of a one-channel photocell. To explore this wavelength dependence, we first compared $\sigma_{II}^2/u_M^2$ and $\Sigma^2/u_M^2$ as a function of input power $u_a$ and $u_b$, and then evaluated $\Sigma^2/u_M^2 = (\sigma_I^2 - \sigma_{II}^2)/u_M^2 = (u_a/u_b u_M^2)(u_b/u_M - 1)$ using $u_i = \int A_i(\lambda) I(\lambda,T) d\lambda$. Figure 3d shows that $\Sigma^2/u_M^2$ is large and positive over the entire range of $u_a$, indicating that two-channel fluctuations $\sigma_{II}$ are reduced below $\sigma_I$. The wavelength dependence in Fig. 2b bottom shows that $\Sigma^2/u_M^2$ is large and positive, except where it reaches a minimum, which corresponds to the spectral peak. For absorption in the green region of the spectrum, fluctuations in the 2QHE photocell become as large as those of a one-channel photocell.

Rather than simply protecting against high light conditions near the spectral peak, natural regulation of energy flow provides tolerance to rapidly changing incident solar power. Under variable solar power, the inputs $u_a$ and $u_b$ will change with $I(\lambda, T)$ (Fig. 4a). If $\Delta u = u_b - u_a$ is small compared to these incident variations, the photocell cannot maintain $u_b > u_M > u_a$ and energy flow is unregulated (Fig. 4a, left). By maximizing $\Delta u$, however, the photocell regulates internal energy flow ($u_b > u_M > u_a$) and will match large external variations to a steady output rate. For general blackbody irradiance, we can approximate $\Delta u$ analytically in order to find the maximum:

$$\Delta u = u_b - u_a \sim |dI(\lambda_0, T_H)/d\lambda| \, F(w, \Delta\lambda/w). \qquad (2)$$

Here, the function $F(w, \Delta\lambda/w)$ integrates over the characteristic absorption line shape of the input channels and is maximized when the peak separation $\Delta\lambda \sim 2\sqrt{2}\, w$ (detailed calculations and analysis are presented in Supplementary Section 3). To maximize equation (2), the center wavelength $\lambda_0$ should lie where the derivative $|dI(\lambda_0, T_H)/d\lambda|$ is largest. At this position, two absorbers with similar input energies $E_a \sim E_b$ will absorb at the largest difference in power $u_b - u_a$.

To illustrate the connection between regulation and the greenness of the quantum photocell, we calculated $\Delta u$ for a blackbody spectrum that shares a similar peak wavelength to that of the measured solar spectrum. The maximum in $\Delta u$ vs. $\lambda_0$ and $\Delta\lambda$ (Fig. 4b) occurs when $\Delta\lambda = 104$ nm for absorption peak of width $w = 35$ nm, in agreement with the analytical result $\Delta\lambda \sim 2\sqrt{2}\,(35\ \text{nm}) \sim 99$ nm (a full comparison is presented in Supplementary Section 4). Similar to the wavelength dependence in Fig. 2b, we observed that $\Delta u$ displays *two* maxima as a function of $\lambda_0$ (Fig. 4a, bottom), with peak values occurring at $\lambda_0 = 335$ nm and 798 nm (Fig. 4b, inset). From $\lambda_0$, $\Delta\lambda$, and $w$, we calculated the differential input energy flux (Fig. 4c) and the corresponding 2QHE energy levels (Fig. 4c, inset). As in Fig. 2b, absorption near the peak ($\lambda_0 \sim 540$ nm) is avoided not because of high irradiance, but instead because the local derivative $|dI(\lambda_0, T_H)/d\lambda|$ vanishes.

**Discussion**

The two-channel QHE structure described here gives rise to a novel regulation mechanism that exhibits compelling similarities to observations in photosynthesis. Several molecular structures have been identified as potential sites for regulation within the minor antenna complexes of photosystem II (PSII)[25-30], and can be directly compared to structures predicted here. By measuring resonant optical excitation (at $\lambda = 650$ nm) and transient kinetics of an individual antenna complex, *Ahn, et al.* provided evidence that excitation energy transfer is regulated by chlorophyll *a* and *b* molecules coupled to individual carotenoids[28]. Further, they proposed that the position of minor antenna complexes, which contain these two-pigment structures, is ideally suited to manipulating the flow of energy from the light-harvesting complex (LHC) to the reaction center. The molecular structure proposed by *Ahn, et al.* strongly resembles the 2QHE photocell predicted by natural regulation (Fig. 2c right), which absorbs near $\lambda = 672$ nm and exhibits strong suppression of power fluctuations (Fig. 3c). Although energy transfer dynamics remain unresolved, recent advances in angstrom-resolution structure determination, such as that of the PSI-LHCI supercomplex[32], may allow for targeted probes that measure the internal power fluctuations and determine exact energy transfer pathways.

Photosynthesis exploits numerous microscopic processes for efficient light energy harvesting that are currently being explored in novel nanoscale photonic technologies[33], yet is unique in its ability to regulate energy flow and store energy in the form of complex organic compounds. Recently, *Kulheim, et al.*[25] measured the lifetime seed production of a highly characterized temperate green plant, *Arabidopsis thaliana*, under natural and synthesized fluctuating light conditions. By comparing plants with and without the proteins necessary for microscopic regulatory feedback, this work established the connection between microscopic regulation processes and macroscopic plant fitness, and highlighted a critical distinction in photosynthesis; Plant fitness is maintained by regulation against variable light conditions rather than protection against high light intensity. This distinction is a key aspect of natural regulation, which links intrinsic regulation of energy flow to an absorption spectrum that closely matches that of photosynthetic green plants.

At present, nanoscale materials that exhibit quantized energy levels are considered to be excellent candidates for light energy harvesting optoelectronics, in part because of their potential for strong absorption, efficient charge and energy transfer, and quantum effects that increase power conversion efficiency[33]. Here, we have quantified power fluctuations in a photocell composed exclusively of quantized energy levels, and determined the energy level structure that mediates intrinsic regulation, an important requirement for solar energy storage. Our findings, together with improved designs that exploit quantum coherence[10-19] to increase power conversion efficiencies beyond theoretical limits[20], make natural regulation a very important step toward ultra-efficient light harvesting devices, and give key insight into the physical origin of the predominance of green plants on Earth.

**Methods**

**Analytical statistical model of power throughput.** Analytical modeling of fluctuating systems out of equilibrium is highly challenging. To overcome this challenge, and in order to model the photocell power throughput, we developed a non-equilibrium statistical model that discretizes time increments and considers the probabilistic absorption or emission of energy at each time step (full details described in Supplementary Section 1). The energy at any time step is a sum of all events in the past, and the energy sequences are then statistical random walks for which the mean and variance of the sum is determined according to the Central Limit Theorem. The steady state behavior, and thus the photocell operation, can be determined from the long-time limit of the stochastic processes.

**Numerical analysis.** We model energy flow in a statistical ensemble of QHEs numerically using standard computational techniques (described in detail in Supplementary Section 2). We include an energy threshold that requires finite energy to enter the system before energy can be emitted from the machine. To examine trends in the calculated power fluctuations, we apply a finite impulse response low-pass filter and iterative search algorithm. We compute $10^2$ energy sequences per computational iteration over a range of probabilities, beginning with the full range of allowed values. The local minimum variance is determined, and the algorithm then selects a narrower range of probabilities around the minimum value to consider in the next iteration. After several subsequent iterations, the algorithm has zoomed in on a small range of probabilities, which includes the true minimum value. In this narrow range, the trend to the minimum value is obscured by local fluctuations and the trend is approximately flat. The minimum value of the two-channel variance is then obtained by averaging the filtered data over this narrow range. Analysis then proceeds by standard methods.

**Acknowledgements**

We thank Alexander Balandin, Chris Bardeen, Veit Elser, Michael Fogler, Nigel Goldenfeld, Tony Heinz, Joseph Iezzi, Paul McEuen, James Sethna, Kyle Shen, Chandra Varma, Feng Wang, and Xiaodong Xu for valuable discussions. This work was supported as part of the SHINES center, an Energy Frontier Research Center funded by the U.S. Department of Energy, Office of Science, Basic Energy Sciences under Award # SC0012670.

**Author Contributions**

N.M.G. and V.A. conceived the research, as well as supervised and analyzed all analytical and numerical work. T.B.A. and Y.B. carried out analytical calculations and numerical simulations. All authors contributed to scientific discussion and to the writing of the manuscript.

Figure 1

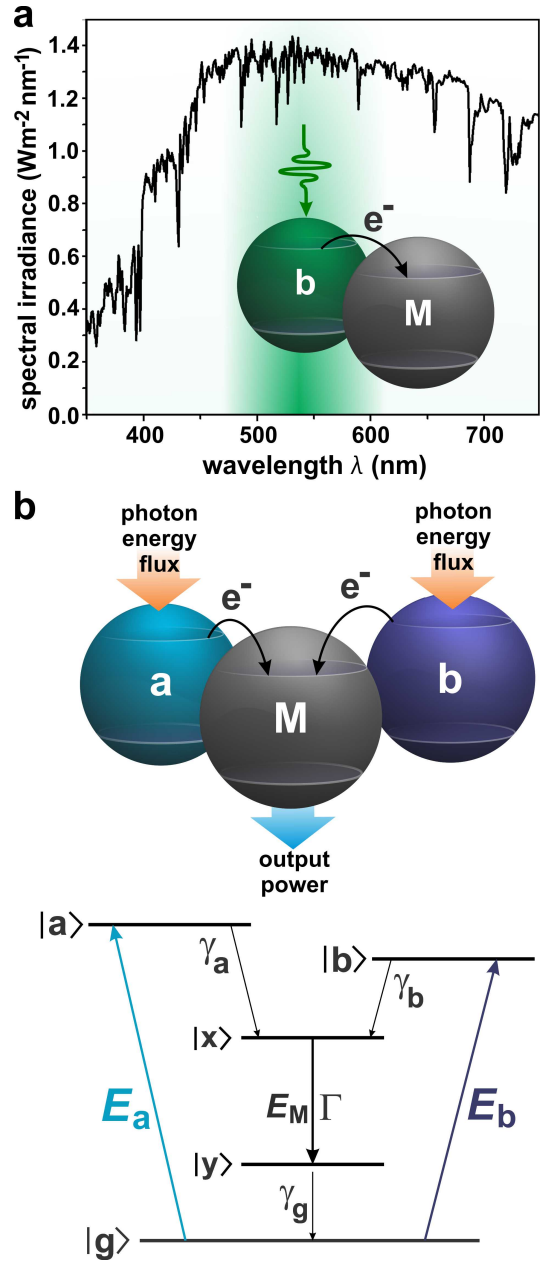

**Figure 1. Solar spectrum and the structure of a two-channel quantum photocell. a,** Solar spectral irradiance measured at the terrestrial surface. Inset: Schematic of a one-channel quantum photocell. **b**, top: Schematic of a two-channel quantum photocell. Photon energy flux enters absorbers *a* and *b*. Output power is generated by the machine *M*. **b**, bottom: Energy level diagram of the 2QHE photocell. $E_a$ and $E_b$ are the excitation energies of the absorbers *a* and *b* with internal transition rates $\gamma_a$ and $\gamma_b$ to the machine *M*. $E_M$ is the output energy characterized by the relaxation rate $\Gamma$ from state $|x\rangle$ to state $|y\rangle$. $\gamma_g$ is the relaxation rate to the ground state $|g\rangle$ of the system.

**Figure 2**

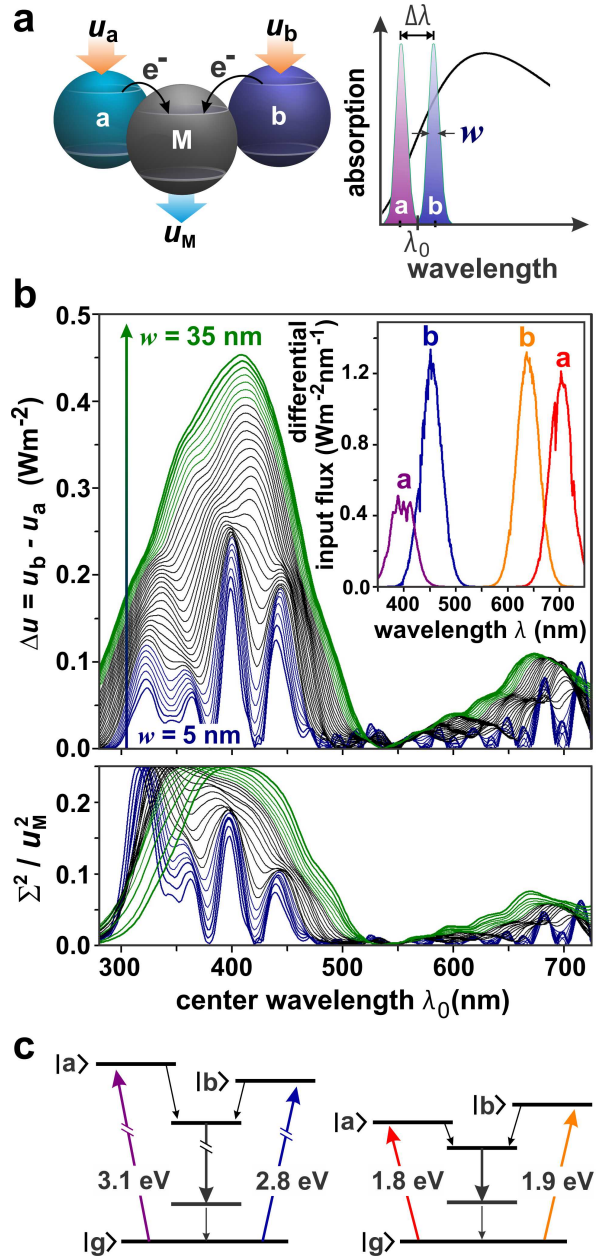

**Figure 2. Quantum structure for natural regulation in a green quantum photocell. a**, left: Power throughput diagram and parameters of the 2QHE photocell: $u_a$ and $u_b$ are the input energy flux (power per unit area) and $u_M$ is the machine output power. **a,** right: Absorption parameters superimposed over a generalized power spectrum. **b**, top: $\Delta u = u_b - u_a$ vs. center wavelength $\lambda_0$, shown for $w$ = 5-35 nm. **b**, top inset: Differential input flux $A_i(\lambda)I(\lambda,T)$ vs. wavelength for $w$ = 35 nm, $\Delta\lambda$ = 102 nm, and $\lambda_0$ = 408 nm (purple, blue) and 672 nm (orange, red). **b**, bottom: $\Sigma^2/u_M^2$ vs. $\lambda_0$ over the same range of $w$ and corresponding $\Delta\lambda$. Several line traces removed for clarity are presented in Supplementary Section 4. **c**, Energy level schematics determined from **b** top inset. Color assignments match corresponding differential energy flux spectra.

Figure 3

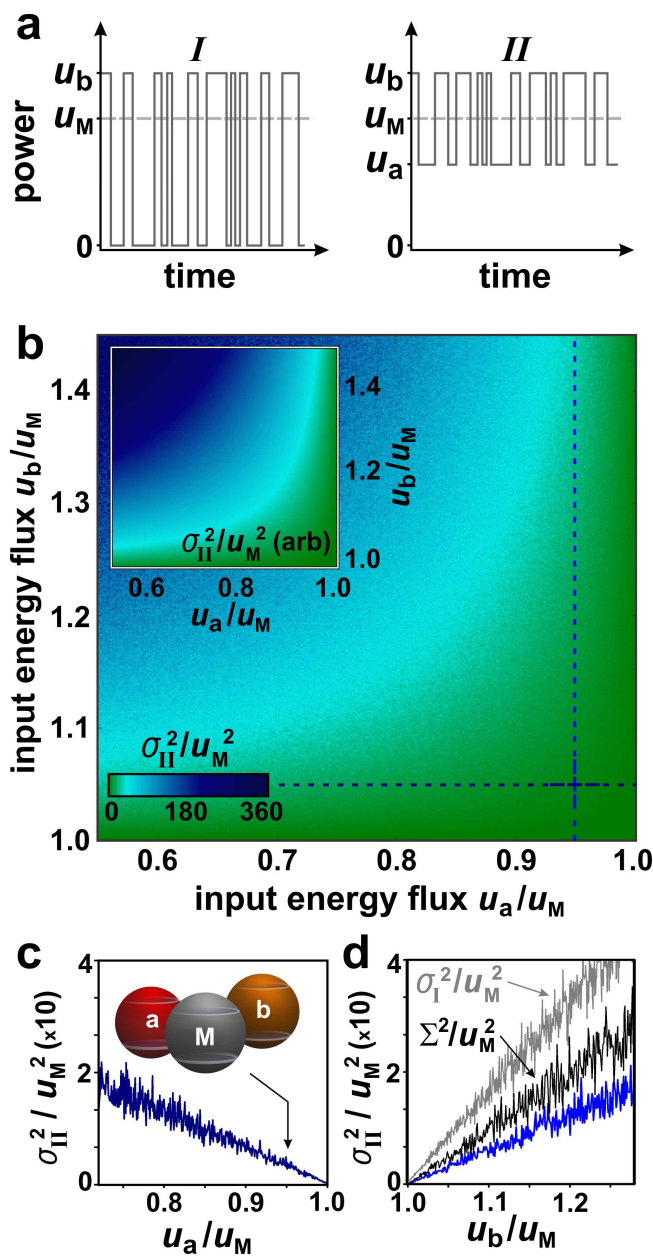

**Figure 3. Suppression of internal power fluctuations in a two-channel quantum photocell.**
**a**, Schematic power modulation in the one (*I*) and two (*II*) channel photocells. **b**, Two-channel variance $\sigma_{II}^2/u_M^2$ vs. $u_a$ and $u_b$, normalized by constant $u_M = (u_b - u_a)/2$. Inset: Analytical dependence of $\sigma_{II}^2/u_M^2$ on $u_a/u_M$ and $u_b/u_M$. **c**, $\sigma_{II}^2/u_M^2$ vs. $u_a/u_M$ for $u_b = 1.05 u_M$ (navy line trace), and **d**, $\sigma_{II}^2/u_M^2$ vs. $u_b/u_M$ for $u_a = 0.95 u_M$ (blue line trace) taken from dashed lines in **b**. The intersection of these lines in **b** marks the variance in the 2QHE centered at $\lambda_0 = 672$ nm (shown schematically in **c**). Panel **d** also shows $\sigma_I^2/u_M^2$ (gray line) and $\Sigma^2/u_M^2$ (black line) vs. $u_b/u_M$.

Figure 4

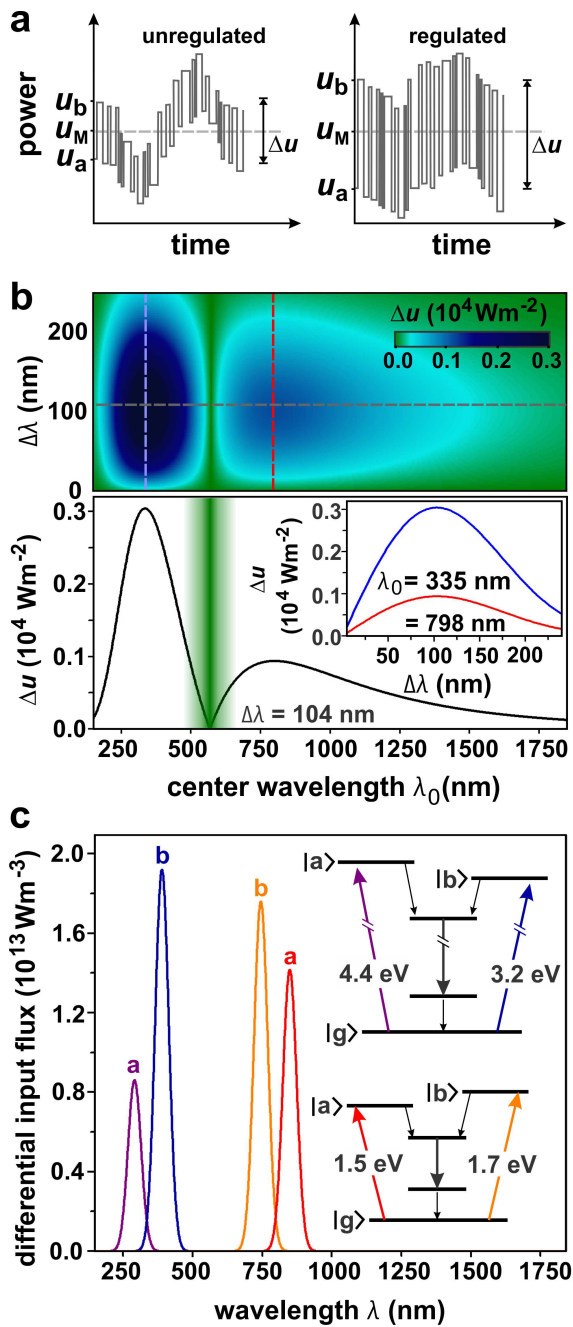

**Figure 4. Natural regulation of energy flow for blackbody irradiance. a**, Schematic power modulation in the two-channel photocell under external irradiance variations. **b**, top: $\Delta u = u_b - u_a$ for a blackbody irradiance at $T_H = 5.5 \times 10^3$ K vs. $\Delta\lambda$ and center wavelength $\lambda_0$. Absorption peak $w = 35$ nm. **b**, bottom: $\Delta u$ vs. $\lambda_0$ at $\Delta\lambda = 104$ nm, taken from the horizontal line trace in **b**, top. Inset: $\Delta u$ vs. $\Delta\lambda$ for $\lambda_0 = 335$ nm (blue) and $\lambda_0 = 798$ nm (red), taken from vertical line traces in **b**, top. Green shaded region shows the minimum $\Delta u$ near $\lambda_0 = 540$ nm. **c**, Differential input flux $A_i(\lambda)I(\lambda,T)$ vs. wavelength for two 2QHEs determined for $w = 35$ nm, $\Delta\lambda = 104$ nm, and $\lambda_0 = 335$ nm (blue, purple) and 798 nm (red, orange). Inset: energy levels determined from **b**. Color assignments match corresponding differential energy flux spectra.

17